\documentclass[superscriptaddress,showpacs,twocolumn]{revtex4}
\def\beq{\begin{equation}}
\def\eeq{\end{equation}}
\usepackage{amsmath}
\usepackage{graphicx}
\usepackage{epsfig}
\usepackage[usenames]{color}

\newcommand{\ket}[1]{| {#1} \rangle}

\usepackage{times}

\begin{document}

\title{Entanglement spectra of critical and near-critical systems in one dimension}

\author{F. Pollmann}
\affiliation{Department of Physics, University of California, Berkeley}
\author{J. E. Moore}
\affiliation{Department of Physics, University of California, Berkeley}
\affiliation{Materials Science Division, Lawrence Berkeley National Laboratory}
\pacs{64.70.Tg, 03.67.Mn,75.10.Pq}

\begin{abstract}
The entanglement spectrum of a pure state of a bipartite system is the full set of eigenvalues of the reduced density matrix obtained from tracing out one part.  Such spectra are known in several cases to contain important information beyond that in the entanglement entropy.  This paper studies the entanglement spectrum for a variety of critical and near-critical quantum lattice models in one dimension, chiefly by the iTEBD numerical method, which enables both integrable and non-integrable models to be studied.
%The results are compared to predictions from conformal mapping and to exact solutions of certain free models.
We find that the distribution of eigenvalues in the entanglement spectra agrees with an approximate result derived by Calabrese and Lefevre~\cite{Calabrese-2008} to an accuracy of a few percent for all models studied.  This result applies whether the correlation length is intrinsic or generated by the finite matrix size accessible in iTEBD.  For the transverse Ising model, the known exact results \cite{Peschel-2009} for the entanglement spectrum are used to confirm the validity of the iTEBD approach.  For more general models, no exact result is available but the iTEBD results directly test the hypothesis that all moments of the reduced density matrix are determined by a single parameter. %%%%%\lookhere%%%%%
\end{abstract}

\maketitle
\section{Introduction}

The past decade has seen a fruitful interplay between problems in condensed matter physics and concepts from quantum information theory.  Entanglement entropy, as reviewed below, has been studied for a wide variety of quantum many-particle systems and has given some considerable insight into both quantum criticality and quantum topological phases.  More recently it has become clear that for some purposes the entanglement entropy is an insufficient characterization of the quantum information in a quantum state.  Both for topological phases and for quantum critical systems, several recent studies have shown that more information about entanglement, the full ``entanglement spectrum'', is required to describe important physical properties.  The purpose of this article is to examine entanglement spectra at and near several quantum critical points in one dimension, chiefly by numerical methods.  This serves as a test of existing theoretical predictions.  The main subject of this paper is a set of numerical tests of the validity of the approximate distribution of eigenvalues in the entanglement spectra derived by Calabrese and Lefevre~\cite{Calabrese-2008}.

The entanglement entropy of a pure state $\ket{\Psi}$ of a bipartite system $AB$ is defined as the von Neumann entropy of reduced density matrix $\rho_A$, obtained by tracing out subsystem $B$.  The result is symmetric with respect to $A$ and $B$:
\beq
S_A = - {\rm Tr}\ \rho_A \log \rho_A = - {\rm Tr}\ \rho_B \log \rho_B = S_B. 
\eeq
This symmetry follows from the Schmidt decomposition, which is convenient for our later discussion.  The Schmidt decomposition \cite{Schmidt-1907} for a bipartite system expresses the original wavefunction as a sum of product states of wavefunctions for the two parts of the system, with the bases chosen for those two parts so that the terms in the sum involve orthogonal wavefunctions:
\begin{equation}
\ket{\Psi}=\sum_{n=1}^{\infty} \lambda_{n} \ket{\Phi_{n A}}\ket{\Phi_{n B}}.
\label{eq:schmidt}
\end{equation}
The $\ket{\Phi_{n A}}$ and $\ket{\Phi_{n B}}$ form orthonormal bases
for the Hilbert spaces of the subsystems  $A$ and $B$, and the $\lambda_{n}$ are positive.
The Schmidt decomposition contains more than one term for entangled states, and the entanglement entropy is
\beq
S=-\sum_{n} \omega_{n} \log \omega_{n}.
\label{eq:defs2}
\eeq
Note that the eigenvalues of the reduced density matrix are the squares of the coefficients appearing in the Schmidt decomposition and $\omega_n=\lambda_n^2$.

Clearly the entanglement entropy is a particular combination of the eigenvalues of the reduced density matrix.  The ``entanglement spectrum'', i.e., the full set of eigenvalues of the reduced density matrix, clearly contains additional information in principle.  For topological phases arising in the quantum Hall effect, the entanglement spectrum has been argued to contain information about topological order~\cite{Li-2008,Regnault-2009} beyond that of the entropy, which contains information about the ``quantum dimensions'' of the topological phase~\cite{Levin-2006,Kitaev-2006}.  The entropy has been a powerful tool in numerical studies to identify topological phases~\cite{Haque-2007,Zosulya-2007}.

The value of entanglement spectrum for critical points, as studied in this paper, is slightly different from its use in measuring topological order.  It has been argued by Calabrese and Lefevre~\cite{Calabrese-2008} that quantum critical points have an entanglement spectrum with a universal one-parameter dependence; since this one parameter can be taken to be the entanglement entropy, one might think that for quantum critical points, the entanglement spectrum contains no more useful information than the entanglement entropy.  However, the detailed form of the entanglement spectrum is important; as an example, the Calabrese-Lefevre approximate form of the entanglement spectrum was used recently as the basis of a theory for ``finite-entanglement scaling'', i.e., how the finite entanglement available in a matrix product state representation affects the observed behavior at quantum critical points.  For this application, understanding not just the entanglement entropy but the full spectrum is essential, and the nature of this spectrum near a one-dimensional quantum critical point is the subject of this paper.

%%%%%\lookhere Might want to advertise conclusions here%%%%

The focus on one-dimensional quantum critical points results from two technical considerations, one numerical and one analytical.  Numerically, the matrix-product-state class of numerical methods (DMRG and its descendants), reviewed in the following section, are especially powerful in one dimension and capable of giving essentially exact ground-state properties for many systems away from criticality.  Analytically, conformal invariance of the $d+1$-dimensional path integral applies at most interesting critical points in any spatial dimension $d$, but only for $d=1$ is the (local) conformal algebra infinite-dimensional.  This infinite-dimensional algebra closely constrains the behavior of one-dimensional systems close to critical points and has led to numerous predictions about entanglement entropy at such critical points, many of which have been confirmed numerically.

In the following section, we review the relevant analytical results on entanglement entropy and entanglement spectrum at one-dimensional critical points.  Section III briefly reviews matrix product states and the Infinite Time-Evolved Block Decimation~\cite{Vidal-2007} (iTEBD) and presents our numerical results on several quantum critical points, including some integrable cases as a check and also some non-integrable cases.

\section{Entanglement entropy and entanglement spectrum}

An important early example of overlap between condensed matter physics and quantum information was the realization that ``most'' quantum critical points in one dimension (those with conformal invariance) have universal entanglement entropy determined by the central charge of the conformal field theory (CFT).  Since central charge was already known to determine properties of the critical point related to entropy, such as the free energy at small nonzero temperature, it is intuitively plausible that the entanglement entropy is also related to this quantity.  For a partition of an infinite one-dimensional chain into a contiguous region of $N$ sites and the infinite remainder, the entanglement entropy scales for large $N$ as~\cite{Holzhey-1994,Vidal-2003,Calabrese-2004}
\beq
S_N \sim {c \over 3} \log N.
\eeq
When the system is away from criticality, this logarithmic divergence is cut off by the correlation length $\xi$, measured in units of the lattice spacing $a$:
\beq
S_N \rightarrow {c \over 3} \log (\xi / a).
\eeq
These entanglement entropies are universal since they depend only on central charge; different critical points in the same universality class have the same entanglement entropy.  An elegant derivation of these results~\cite{Calabrese-2004} is via a replica trick that obtains the entanglement entropy using the theory defined on a particular Riemann surface; the central charge then appears via the mapping of an ordinary plane to this Riemann surface.  It is believed that the full entanglement spectrum that leads to this entropy is not universal but nearly so~\cite{Calabrese-2008}, as now explained.  Since the entanglement spectrum is not universal, it should not follow from the conformal mapping method that makes clear the universality of the entropy.

There are only a few results on the behavior of entanglement entropy at translation-invariant critical points above one spatial dimension~ \cite{Ryu-2006, Fradkin-2006,Metlitski-2009,Hsu-2009,Furukawa-2009,Wolf-2006,Srednicki-1993,Cramer-2006} and in general the entanglement spectrum has not been considered.  Hence we focus on the entanglement spectrum at or near a one-dimensional quantum critical point with conformal invariance.  The entanglement spectrum can be obtained from traces of all powers of the reduced density matrix:
\beq
R_\alpha \equiv {\rm Tr} \rho_A^\alpha.
\eeq
In order to make the entanglement spectrum well-defined, we have to introduce a length scale $L_{\rm eff}$ coming from either a block size or a correlation length, just as for the entanglement entropy.
When this length scale is much larger than any microscopic scale, a scaling analysis~\cite{Calabrese-2004} predicts
\beq
R_\alpha = c_\alpha L_{\rm eff}^{-c (\alpha - 1/\alpha)/6},
\eeq
where $c_\alpha$ are nonuniversal constants.  Following Ref.~\onlinecite{Calabrese-2008}, we introduce the parameter $b$ and rewrite this as
\beq
R_\alpha = c_\alpha e^{-b (\alpha - 1/\alpha)},\quad b = {c \over 6} \log L_{\rm eff}
\eeq
and can obtain the form of the entanglement spectrum immediately {\it under the assumption} that the $c_\alpha$ are constant.  This assumption of constant $c_\alpha$ was shown to be a good approximation for the $XX$ model and even on the whole $XXZ$ line (more details are given later).  However, it seems valuable to evaluate the validity of this assumption more generally, including for models that are not integrable.  Note that the parameter $b$ is simply related to the entanglement entropy $S$ obtained for this distribution: $b = S/2$.  For conformally invariant models, $b$ is related to the largest eigenvalue of the density matrix  $b=-\log\omega_{\rm max}$. This is the so called single-copy entanglement \cite{Eisert-2005, Peschel-2005, Orus-2006}. We found numerically, that this relation holds in good approximation if we slightly detune the system from criticality.

Under the assumption of constant $c_\alpha$, a Laplace transform leads to a universal one-parameter distribution of density matrix eigenvalues near the critical point~\cite{Calabrese-2008}.  The parameter appearing in the distribution, which we take to be $b$, combines both an ``intrinsic'' property of the critical point (the central charge) and an ``extrinsic'' one (either the correlation length or the block size, whichever is smaller).  An intuitive way to express this distribution is via the mean number of eigenvalues larger than a given value $\omega$,
\beq
n(\omega) = \int_\omega^{\omega_{\rm max}}\,d\omega\,P(\omega) = I_0(2 \sqrt{b \log(\omega_{\rm max}/\omega)})\label{eq:lev_for}.
\eeq
Here the largest eigenvalue $\omega_{\rm max}$ is determined by the one parameter $b$: $b = - \log \omega_{\rm max}$.  As $b$ increases, the tail of the distribution becomes longer.  Examples of this distribution are shown in FIGS.~\ref{fig:isi_lev}--\ref{fig:spin1_lev} below.  The entanglement spectrum of 1D systems away from criticality has also been discussed in the context of majorization theory in Ref.~\onlinecite{Orus-2005}.
%%%%%\lookhere%%%%%%%%%% 

%\lookhere Peschel results
% for free lattice systems~\cite{peschelreview} 
\section{Numerical methods and results}

\subsection{Methods}

Our main results are obtained using the Infinite Time-Evolving Block Decimation (iTEBD) algorithm \cite{Vidal-2007} to study several one-dimensional Hamiltonians with translationally invariant ground states.  
This algorithm can be viewed as a descendant of the density matrix renormalization group (DMRG) algorithm \cite{White-1992}. Both DMRG and iTEBD construct trial wavefunctions that are ``matrix product states'' (MPS) \cite{Ostlund-1995}.  We review the basics of such states in order to understand their connection to the entanglement spectrum.

We consider a system with $N$ sites and periodic boundary conditions, where each site has $d$ orthogonal states.  Any pure state of the system is a superposition of the product basis states $\ket{s_1s_2\dots s_N}=\ket{\{s\}}$ where $1\leq s_i\leq d$.  A MPS for such a system has the form
\begin{equation}
|\psi\rangle=\sum^d_{s_1,\dots, s_{N}=1}\mathrm{Tr}\left[{A^{[1]}_{s_1}\dots A^{[N]}_{s_{N}}} \right]|s_1\rangle\dots|s_{N}\rangle.
\label{eq:mps}
\end{equation}
For each site $i$ there are $d$ matrices $A^{i}_{s_i}$ of a finite dimension $\chi \times \chi$. A product (unentangled) wavefunction for a chain of particles/spins is obtained by multiplying together scalar amplitudes for the particle/spin state at each site.  The MPS generates entanglement by using matrices for each site and can store an increasing amount of information as the matrix dimension $\chi$ increases.  For fixed $\chi$, there are at most $\chi$ nonzero eigenvalues of the reduced density matrix, and hence the maximum possible entanglement is $\log \chi$.

The advantage of requiring translational invariance is that there are only finitely many different matrices $A^i_{s_i}$.  The iTEBD algorithm finds an approximation of the ground state by performing an imaginary time evolution of a randomly chosen initial MPS $|\psi^{\chi}_{\text{init}}\rangle$ with fixed dimension $\chi$. It is therefore required that $|\psi^{\chi}_{\text{init}}\rangle$  has a non-zero overlap with the ground state.  Because this method always constructs wavefunctions for the infinite system, its errors result from finite entanglement rather than finite size, and one application of the entanglement spectrum was to develop a theory~\cite{Pollmann-2009} for the ``finite-entanglement scaling'' of resulting errors~\cite{Tagliacozzo-2008}: the effective correlation length induced by finite entanglement for a critical Hamiltonian scales as $\xi \propto \chi^\kappa$, where the exponent $\kappa$ is determined by the central charge at the critical point. For the iTEBD simulations presented in this paper, we used a second order Trotter decomposition for the imaginary time evolution with decreasing time steps $\delta t=10^{-1},\dots,10^{-8}$. We perform the evolution until the energy is converged up to 12 digits and the entanglement entropy up to 6 digits. The run times of the algorithm depends very strongly on the model and the parameters used \cite{Vidal-2007}. 

We will limit ourselves in this paper to cases in which the iTEBD algorithm converges effectively to the matrices of a given dimension that are the best approximation of the ground state.  In a gapped phase, the entanglement entropy of the exact wavefunction is finite, and the approximation by MPS's converges rapidly~\cite{Eisert-2008}.  The iTEBD result should therefore converge to the physically correct distribution if the correlation length $\xi$ (i.e., the correlation length in the physical system, not the finite-entanglement approximation) satisfies
\beq
a \ll \xi \ll \log \chi,
\eeq
where $a$ is the correlation length and $\chi$ is the matrix size.  When these conditions are satisfied, the iTEBD result is not evolving rapidly with $\chi$, and the numerical spectrum should be that of the physical system.  The approach of the numerical spectrum to the physical one is somewhat subtle, because the physical spectrum even at finite correlation length has an infinite number of nonzero eigenvalues, while the numerical spectrum has at most $\chi$.  However, the fraction of total probability in the tail of small eigenvalues becomes very small with increasing $\chi$ away from criticality, unlike at the critical point.

Note that this limit is different from the finite-entanglement scaling limit $\log \chi \ll \xi$, where the physical correlation length is effectively infinite but there is an apparent correlation length in numerics.  Hence a different type of result is expected when the correlation length satisfies $\xi \gg \log \chi$.  It was observed previously~\cite{Pollmann-2009} that the Calabrese-Lefevre distribution approximately describes the spectra obtained in this limit, where the correlation length is generated by finite $\chi$.  In the following section, we compare numerical spectra to the Calabrese-Lefevre distribution and in some cases to exact results.

\section{Numerical results}
\subsection{Transverse Ising model}

\begin{figure}[hbp]
\begin{center}
\includegraphics[width=85mm]{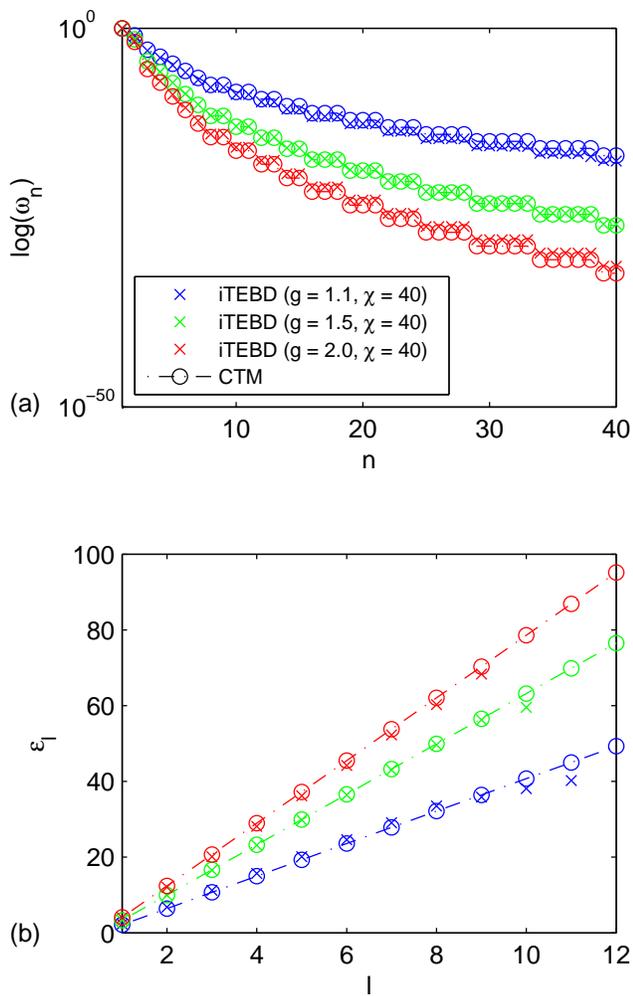}
\end{center}
\caption{Entanglement (Schmidt) spectra of the transverse field Ising model for different parameter.  Panel (a) compares the numerical iTEBD-results ($\times$) with the entanglement spectrum obtained by the CTM approach ($\circ$) \cite{Peschel-2009,Peschel-2004}. Panel (b) compares directly the single particle eigenvalues.\label{fig:isi_spe}}
\end{figure}

\begin{figure}[hbp]
\begin{center}
\includegraphics[width=85mm]{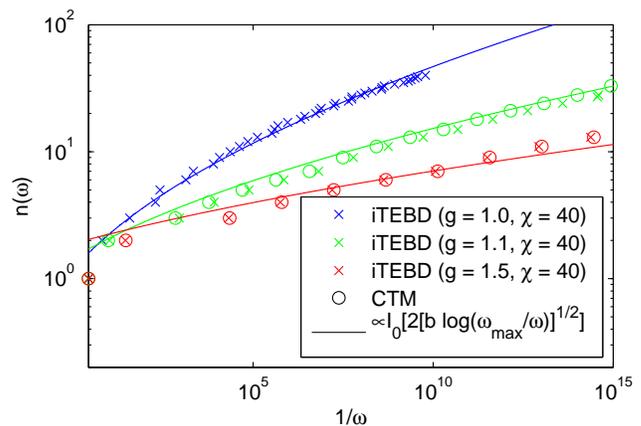}
\end{center}
\caption{Comparison the prediction by the formula $n(\lambda)  = I_0(2 \sqrt{b \log(\lambda_{\rm max}/\lambda)})$ with the results obtained by the iTEBD algorithm and the CTM approach for the transverse Ising model near the critical point at $g=1$.\label{fig:isi_lev}}
\end{figure}

We start by considering the quantum Ising model whose Hamiltonian is
\begin{equation}\nonumber
H = -\sum_i \left(\sigma_i^x \sigma_{i+1}^x + g \sigma_i^z\right),
\label{Eq:hamising}
\end{equation}
with the Pauli matrices  $\sigma_i^{\alpha}$ and $g\geq0$. The model is critical at $g=1$ with central charge $c=1/2$. This model belongs to a class for which, the entanglement spectrum can be calculated by exploiting the relation between quantum chains and two-dimensional classical models \cite{Peschel-2009, Peschel-1999, Peschel-2004}. Using a corner transfer matrix approach, it can be show that the reduced density matrix has a diagonal form
\begin{equation}
\rho=\mathcal{K}\exp\left( -\sum_{j\geq0}\epsilon_jc_j^{\dag}c^{\vphantom{\dag}}_j\right)\label{rho_peschel}
\end{equation} 
with fermionic operators $c_j^{\dag},\ c^{\vphantom{\dag}}_j$ and single particle eigenvalues
\begin{equation}
\epsilon_j=\begin{cases}
(2l+1)\epsilon&, g>1\\
\ 2l\epsilon&, g<1	
\end{cases}
\end{equation}
where $l=0,1,2,\dots$. The value of $\epsilon$ is given by
\begin{equation}
\epsilon=\pi I(k^{\prime})/I(k)
\end{equation}
where $I(k)$ is the complete elliptic integral of the first kind and $k'=\sqrt{1-k^2}$. The parameter  $k$, is for  the case of the transverse Ising model, simply given by
\begin{equation}
k=\begin{cases}
1/g&,g>1\\
g&,g<1.
\end{cases}
\end{equation}
We compare the eigenspectrum $\omega_n$ of the reduced density matrix $\rho$ in Eqn.~(\ref{rho_peschel}) with our numerical results.  FIG.~\ref{fig:isi_spe}(a) shows the good agreement of the entanglement spectra obtained from Eqn.~(\ref{rho_peschel}) and the numerical results which we obtained using the iTEBD algorithm. FIG.~\ref{fig:isi_spe}(b) shows single particle eigenvalues and compares the analytic results from  Eqn.~(\ref{rho_peschel})  with the numerical results. In particular, we observe a linear dispersion which is predicted by the CTM approach. If we approach the critical point, the value of $\chi$ has to be chosen increasingly large to get a good agreement.   Our findings agree with previous numerical studies for these models in Refs.~\onlinecite{Vidal-2007, Peschel-1999, Orus-2005}.

Next we calculate the distribution function $n(\omega)$ using the iTEBD algorithm and the CTM approach. The results are shown  in FIG.~\ref{fig:isi_lev}.  At the critical point (g=1), the correlation length $\xi$ diverges and the  actual distribution of the eigenvalues is flat. The iTEBD algorithm, however, truncates the entanglement spectrum, thereby generating a finite correlation length. Thus the entanglement entropy is finite \cite{Pollmann-2009} and we use the resulting $b=S/2$ as a parameter in $n(\omega)$. If we detune the system from criticality, the correlation length is finite and the iTEBD is essentially exact for a $\chi$ which is large enough. Consequently, the iTEBD algorithm and the CTM approach give the same results. The distribution function $n(\alpha)$ in Eqn.~(\ref{eq:lev_for})  agrees well with the iTEBD results, independent of whether the correlation length is induced by the truncation or an actual finite correlation length. If the systems is strongly detuned from criticality, the discrete structure of the entanglement spectrum -- which is not captured by Eqn.~(\ref{eq:lev_for}) -- becomes more pronounced. 
\subsection{XXZ model}
\begin{figure}[htp]
\begin{center}
\includegraphics[width=85mm]{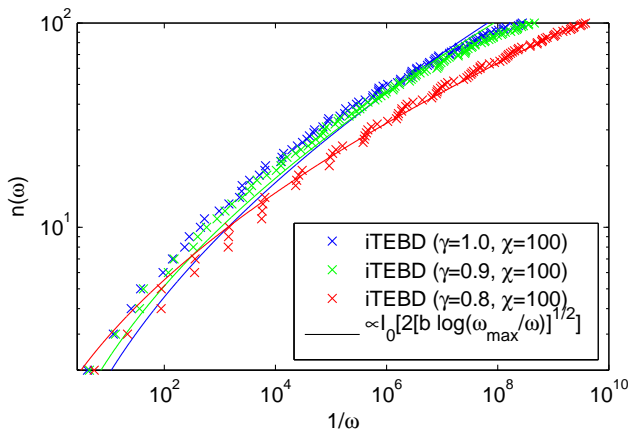}
\end{center}
\caption{Comparison the prediction by the formula $n(\omega)  = I_0(2 \sqrt{b \log(\omega_{\rm max}/\omega)})$ with the results obtained by the iTEBD algorithm  for the XXZ model near the critical point at $\gamma=1$.\label{fig:xxz_lev}}
\end{figure}
The $XXZ$ spin chain is described by the Hamiltonian
\begin{equation}\nonumber
H = \sum_i \left(\sigma_i^x \sigma_{i+1}^x + \sigma_i^y \sigma_{i+1}^y + \gamma \sigma_i^z\sigma_{i+1}^z\right)
\label{Eq:hamxxz}
\end{equation}
and is critical in the entire range $\gamma \in [0,1]$ with central charge $c=1$. Critical exponents change continuously with $\gamma$ \cite{Baxter-1982}.
The results for the distribution function $n(\omega)$ near the Heisenberg (XXX) point at $\gamma=1$ are shown  in FIG.~\ref{fig:xxz_lev}. For all parameters, the distribution functions Eqn.~(\ref{eq:lev_for})  describes the iTEBD results well. A detailed study of the entanglement spectrum of XXZ chains can be found in Ref.~\onlinecite{Nienhuis-2009}.

\subsection{Spin-1 model}
\begin{figure}[htp]
\begin{center}
\includegraphics[width=85mm]{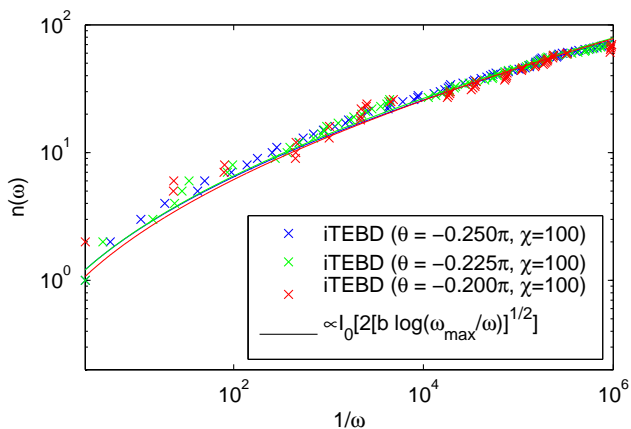}
\end{center}
\caption{Comparison the prediction by the formula $n(\omega)  = I_0(2 \sqrt{b \log(\omega_{\rm max}/\omega)})$ with the results obtained by the iTEBD algorithm  for the spin-1 model near the critical point at $\theta=-\pi/4$ \label{fig:spin1_lev}}
\end{figure}

The two models we have studied so far are integrable models. Now we consider  the $S=1$ Heisenberg chain with biquadratic term,
\begin{equation}\nonumber
H = \sum_i \left[\cos \theta ({\bf S}_i {\bf .} {\bf S}_{i+1}) + \sin \theta ({\bf S}_i {\bf .} {\bf S}_{i+1})^2\right],
\label{eq:hamspin1}
\end{equation}
which is generally non-integrable. This system is known to have two exactly solvable critical points, the SU$(2)_2$ point at $\theta=-\pi/4$ with $c=3/2$ \cite{Alcaraz-1988} and the SU$(3)_1$ point at $\theta=\pi/4$ with $c=2$ \cite{Itoi-1997}. The entire region $\theta \in [\pi/4, \pi/2)$ is critical with $c=2$ \cite{Lauchli-2006}. A detailed study of critical SU($N$) can be found in Ref.~\cite{Fuehringer-2008}. The distribution functions $n(\omega)$ near the critical point at $\theta=-\pi/4$ are shown in FIG.~\ref{fig:spin1_lev}. For all parameters, the distribution functions Eqn.~(\ref{eq:lev_for})  describes the iTEBD results well.

\subsection{Moments of the reduced density matrix}
\begin{figure}[ht]
\begin{center}
\includegraphics[width=85mm]{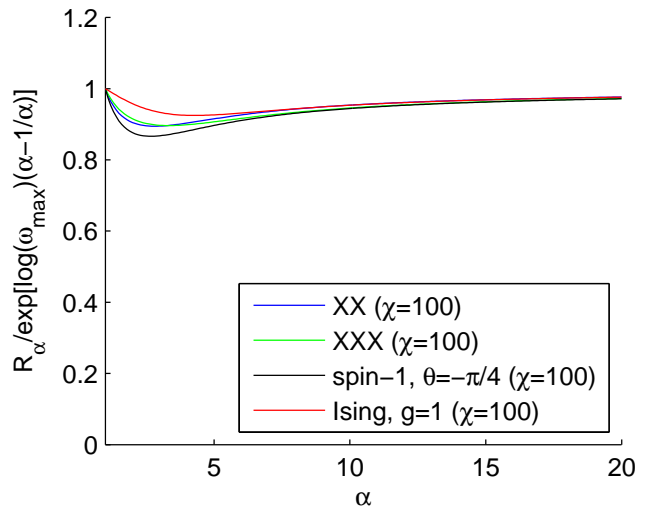}
\end{center}
\caption{The quantity $c_{\alpha}=R_{\alpha}/ e^{-b (\alpha - 1/\alpha)}$ with $b=-\log\omega_{\text{max}}$ calculated numerically for critical points of different models.}
\label{moments}
\end{figure}
%%%%%%%%%%%%%%%%%
The derivation of the distribution function $n(\omega)$ in Eqn.~(\ref{eq:lev_for}) has been made under the assumption that the moments of the reduced density can be expressed as $R_\alpha = c_\alpha e^{-b (\alpha - 1/\alpha)}$ with $c_{\alpha}$ being constant. In order to check the validity of this assumption, we calculated the moments $R_{\alpha}=\text{Tr}\rho^{\alpha}$ explicitly using the iTEBD algorithm. The results for $c_{\alpha}=R_{\alpha}/ e^{-b (\alpha - 1/\alpha)}$ and $\chi=100$  are shown in FIG.~\ref{moments}. We find $c_{\alpha}=1$ in two cases: (i) If $\alpha=1$ because $\text{Tr}\rho=1$  (ii) For $\alpha\rightarrow\infty$ because in this case,  $R_{\alpha}=\omega_{\text{max}}^{\alpha}=e^{-\alpha b}$. In between, the $c_{\alpha}$ deviate from unity by at most $~10\%$. This explains the good agreements of the iTEBD results and the distribution function in Eqn.~(\ref{eq:lev_for}).      
%%%%%%%%%%%%%%%%

\section{summary}
For all considered models, the distribution function $n(\omega)$ derived by Calabrese and Lefevre agrees at the few percent level with the iTEBD results, independent of whether the correlation length is induced by the truncation or an actual finite correlation length.  A direct calculation of the  moments $c_\alpha$ suggests that the assumption  $c_\alpha=\text{const.}$ is a good approximation.  For future work, it would be interesting to understand theoretically why this is the case.  In general, one would expect nonuniversal quantities such as the $c_\alpha$ not to have any constraint on their variation, in which case it is puzzling why the standard models we have studied all have roughly constant $c_\alpha$.  Conversely, if there are actually bounds on the $c_\alpha$ that explain the observed behavior (that they are not constant but nearly so), then understanding the origin of those bounds is an important problem.  It is interesting to note in FIG.~\ref{moments} that the model with the strongest deviations from the constant behavior of moments $c_\alpha$ is the spin-1 model, which is the most complicated analytically and also the most difficult numerically (in part because of its higher central charge, which means that more degrees of freedom are critical).

The results of this paper show that the Calabrese-Lefevre distribution is indeed a good starting point, at least for a number of integrable and non-integrable models, for applications of the entanglement spectrum at critical points.  They also place constraints on the ultimate theory of this entanglement spectrum including non-universal effects.  The authors acknowledge support from ARO (F.~P.) and NSF DMR-0804413 (J.~E.~M.).

%\bibliographystyle{apsrev1}
%\bibliography{bib_entanglement}
\end{document}